%% file: morello_ifae2011.tex
      [1999/12/01 v1.4c Il Nuovo Cimento]
\documentclass{cimento}

\input{macro}

\newcommand{\betas}{\ensuremath{\beta_{s}}}   
\newcommand{\betasSM}{\ensuremath{\beta_{s}^{\mathrm{SM}}}}	
\newcommand{\betasNP}{\ensuremath{\beta_{s}^{\mathrm{NP}}}}	

\newcommand{\phis}{\ensuremath{\phi_{s}}}   
\newcommand{\phisSM}{\ensuremath{\phi_{s}^{\mathrm{SM}}}}

\newcommand{\DGSM}{\ensuremath{\Delta\Gamma^{\mathrm{SM}}}}

\usepackage{graphicx}  
\title{B Physics at the TeVatron}
\author{M.J.~Morello\from{ins:x}\thanks{for the CDF and D\O\ Collaborations}}
\instlist{\inst{ins:x} University of Pisa \atque INFN of Pisa - Largo
  Bruno Pontecorvo 3, 56127 Pisa, Italy}
\PACSes{\PACSit{13.20.He}{Decays of bottom hadrons}
\PACSit{13.25.Ft}{Decays of charmed hadrons}
\PACSit{13.30.Eg}{Hadronic decays}}
\begin{document}

\maketitle

\begin{abstract}
The CDF and D\O\ experiments at the Tevatron $p\bar{p}$ collider established that extensive and detailed exploration of the $b$--quark 
dynamics is possible in hadron collisions, with results competitive and supplementary to those from $e^+e^-$ colliders. 
This provides a rich, and highly rewarding program that is currently reaching full maturity. 
I report a few recent world-leading results on rare decays, CP-violation in $B^0_s$ mixing, $b\to s$ penguin decays, and charm physics.
\end{abstract}

\section{Introduction}

Precise results from the successful $B$~factory experiments disfavor large contributions from 
non-standard model (SM) physics in tree-dominated bottom meson decays. Agreement with the 
SM within theory uncertainties is also manifest in higher-order processes, such as 
$K^0$--$\bar{K}^0$ or $B^0$--$\bar{B}^0$ flavor mixing. The emerging picture confirms the 
Cabibbo-Kobayashi-Maskawa (CKM) framework as the leading pattern of flavor dynamics.  
Non-SM contributions, if any,  are small corrections, or appear well beyond the TeV scale (and the LHC reach), 
or have an unnatural, highly fine-tuned flavor structure that escaped all experimental tests to date. 
The last chances to avoid such a disappointing impasse include the extensive study of the 
physics of the bottom-strange mesons, still fairly unexplored, along with a few rare $B^0$ decays, not fully probed at the $B$~factories 
because of limited event statistics.

CDF and D\O\ experiments at the Tevatron are currently leading the exploration of this physics,  
owing to CP-symmetric initial states in $\sqrt{s}=1.96$ TeV $p\bar{p}$ collisions, 
large event samples collected by well-understood detectors, and mature analysis techniques.  
CDF and D\O\ have currently collected more than 8~\lumifb\ of physics-quality data. 
The sample size will reach 10~\lumifb\ in October 2011. 

In the following I report some recent, world-leading results,  selected among those more sensitive to the presence of non-SM particles 
or couplings. Branching fractions indicate \CP-averages,  
charge-conjugate decays are implied everywhere.

\section{Rare $B \to\mu^+\mu^-$ and  $B\to h\mu^+\mu^-$ decays}

Decays mediated by flavor changing neutral currents,  such as $B^0_{(s)} \to \mu^+\mu^-$ or  $B\to h\mu^+\mu^-$ are highly suppressed in the 
SM because they occur only through higher order loop 
diagrams. Their phenomenology provide enhanced sensitivity to a broad class of non-SM contributions.

The $B^0_{(s)} \to \mu^+\mu^-$  rate is proportional to the CKM matrix element $|V_{td}|^2 (|V_{ts}|^2)$, and is further suppressed by helicity factors.
 The SM expectations for these branching fractions are $\mathcal{O}(10^{-9})$,  ten times smaller than the current experimental sensitivity.  
An observation of these decays at the Tevatron would unambiguously indicate physics beyond the SM. 
On the other hand, improved exclusion-limits 
strongly constrain the available space of parameters of several models of supersymmetry~(SUSY).
The latest CDF search for  $B^0_{(s)}\to \mu^+\mu^-$ decays uses  3.7~ \lumifb.
The resulting 90 (95)\% CL upper-limits are $\br(B^0_s \to \mu^+\mu^-) < 3.6 (4.3)\times 10^{-8}$ and $\br(B^0 \to \mu^+\mu^-) < 6.0 (7.6)\times 10^{-9}$~\cite{bsmumu}.
These results are the most stringent currently available and reduce  significantly the allowed parameter space for a broad range of SUSY models. 
An update of this analysis with approximately double sized sample will further improve them soon. The CDF expected 95\%~CL upper-limit
is $\br(B^0_s \to \mu^+\mu^-) < 2 \times 10^{-8}$~\cite{walter_beauty},  on a data sample corresponding to 
approximatively 7~\lumifb. D\O\ performed a similar analysis using a data sample of about 6.1~\lumifb. 
The resulting 90 (95)\% CL upper-limit is $\br(B^0_s \to \mu^+\mu^-) < 4.2 (5.1)\times 10^{-8}$~\cite{d0_bsmumu}.

CDF also  updated the analysis of $B^0\to K^{*}(892)^{0}\mu^+\mu^-$, $B^+\to K^+\mu^+\mu^-$, 
and $B^0_s\to\phi\mu^+\mu^-$ decays to 4.4~\lumifb of data~\cite{afb}.  These are suppressed in the SM ($\br\approx 10^{-6}$), 
with amplitudes dominated by penguin and box $b\to s$ transitions.  Despite the presence of final-state hadrons, 
accurate predictions greatly sensitive to non-SM contributions are possible for relative quantities based on angular-distributions 
of final state particles~\cite{afb_th}. 
Prominent signals of $120\pm16$ $B^+\to K^+\mu^+\mu^-$ and $101\pm12$ $B^0\to K^{*0}\mu^+\mu^-$ events   
are observed. The absolute branching fractions, measured using the resonant decays as a reference,  
are $[0.38 \pm 0.05\stat \pm 0.03\syst] \times 10^{-6}$ and $[1.06 \pm 0.14\stat \pm 0.09\syst] \times 10^{-6}$, 
respectively, consistent and competitive with previous determinations~\cita{pdg}. In addition, $27\pm 6$ $B^0_s\to\phi\mu^+\mu^-$ events are reconstructed, 
corresponding to the first observation of this decay, with a significance in excess of $6\sigma$.  
The branching ratio, $[1.44\pm 0.33\stat \pm 0.46\syst]\times 10^{-5}$, is consistent with theoretical prediction, 
and corresponds to the rarest \bs\ decay ever observed to date.

\section{Measurement of the \bs\ mixing phase}
Non-SM contributions have not yet been excluded in \bs-\abs\ mixing. Their magnitude is constrained to be small 
by the precise determination of the  frequency \cita{mixing}. However, knowledge of only the frequency leaves possible non-SM 
contributions to the unconstrained (\CP-violating) mixing phase. The time evolution of  flavor-tagged \bsjpsiphi\ decays allows
 a determination of this phase largely free from theoretical uncertainties. These decays probe the phase-difference between the mixing 
and the $\bar{b}\to \bar{c}c\bar{s}$ quark-level transition, $\betas = \betasSM+\betasNP$, which equals $\betasSM=\arg(-V_{ts}V_{tb}^{*}/V_{cs}V_{cb}^{*}) \approx 0.02$ 
in the SM  and is extremely sensitive to non-SM physics in the mixing. A non-SM contribution  ($\betasNP$)  would also enter $\phis = \phisSM - 2\betasNP$, 
which is the phase difference between direct decay and decay via mixing
into final states common to \bs\ and \abs, and is also tiny in the SM: 
$\phisSM = \arg(-M_{12}/\Gamma_{12}) \approx 0.004$. Because the SM values for \betas\ and \phis\ cannot be resolved with the 
precision of current experiments, the following approximation is used: $\phis \approx -2\betasNP \approx -2\betas$, which 
holds in case of sizable non-SM contributions. Note that the phase \phis\ also modifies the decay-width difference between light and 
heavy states, $\Delta\Gamma=\Gamma_L-\Gamma_H=2|\Gamma_{12}|\cos(\phis)$, which enters in the \bsjpsiphi\ amplitude 
and equals $\DGSM \approx 2|\Gamma_{12}| = 0.086 \pm 0.025$ ps$^{-1}$ in the SM \cita{nierste}.

CDF and D\O\ updated the measurement of the time-evolution of flavor-tagged 
\bsjpsiphi\ decays respectively to a sample of 5.2~\lumifb\ and 6.1~\lumifb~\cite{sin2betas,sin2betas_d0}. 
\begin{figure}[hb]
\centering
\includegraphics[height=4.4cm,angle=0]{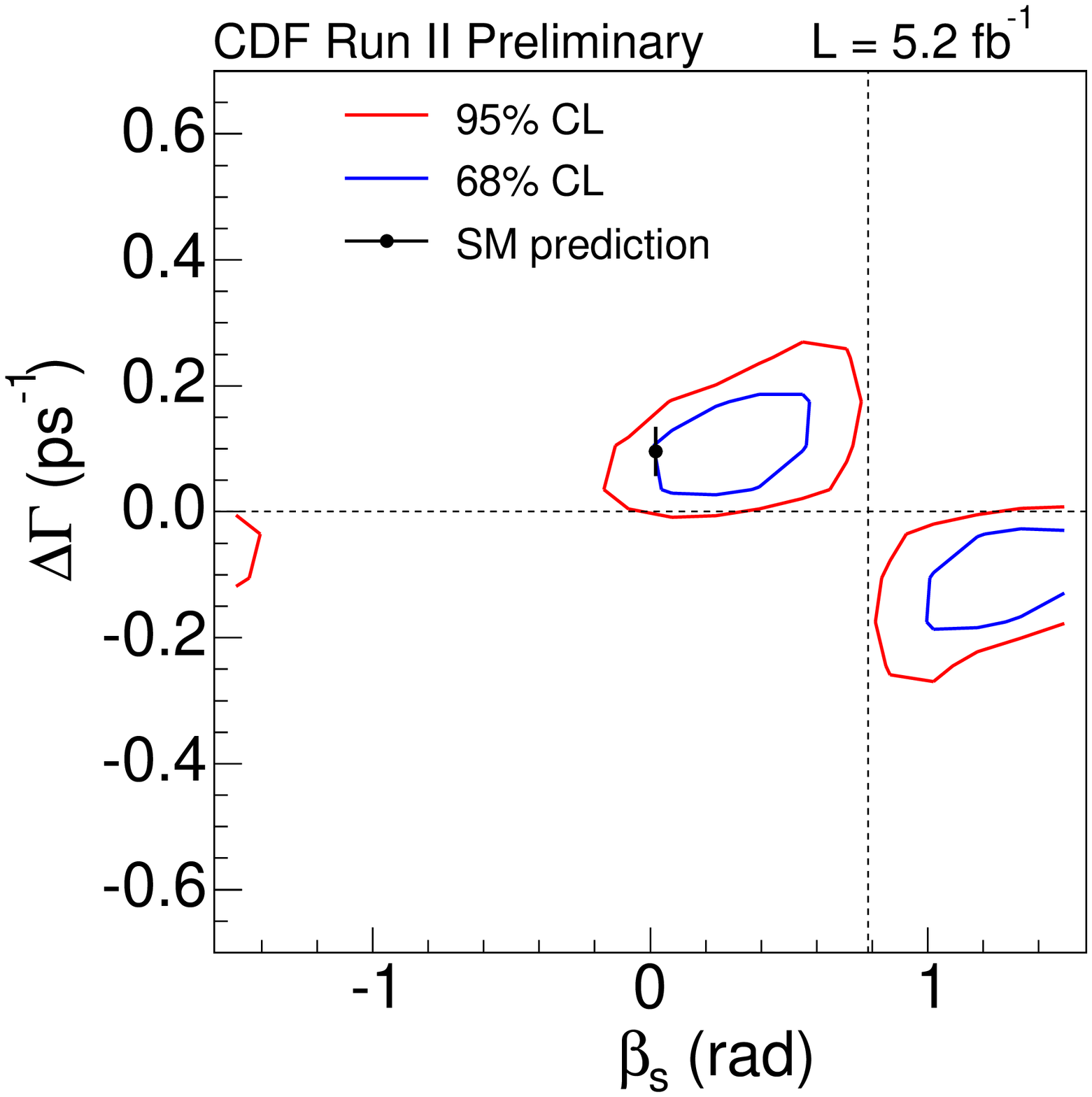}
\hspace{0.5cm}
\includegraphics[height=4.4cm,angle=0]{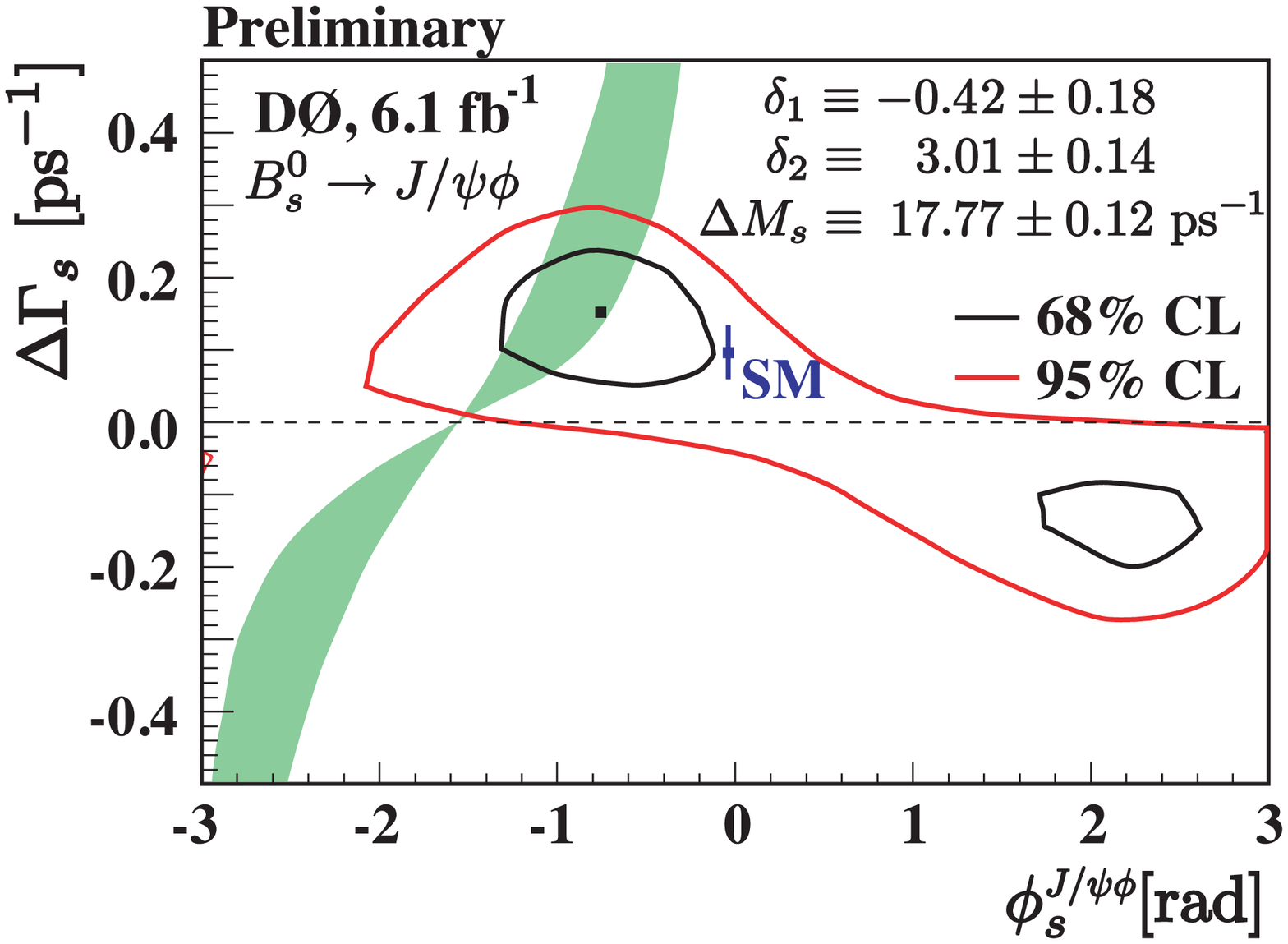}
\caption{\label{fig:contours}Confidence region in the $(\betas,\Delta\Gamma)$ plane for CDF (left pannel), in the $(\phi_{s},\Delta\Gamma)$ plane for D\O\ (right pannel).}
\end{figure}
With phase floating the resulting allowed regions in the ($\beta_s,$ $\Delta\Gamma_s$) plane  are shown in Fig.~\ref{fig:contours}.
They are  greatly reduced with respect to the previous measurements~\cite{betas_old}.
CDF is fairly consistent  with the SM, at $0.8\sigma$ level with no external constraints. D\O\ reports a similar result, 
but using external inputs for strong phases $\delta_1=-0.42\pm0.18$ and $\delta_2=3.01 \pm 0.14$~\cite{pdg}
and $\Delta M_s = 17.77 \pm 0.12$~ps$^{-1}$~\cite{mixing}.

The decay \bsjpsiphi\ has a mixture of the CP-even and -odd components in the final state
and an angular analysis is needed to separate them.
A sufficiently copious \bsjpsifzero\ signal with $f_{0}(980)\to \pi^+\pi^-$, and tagged \bs\ production flavor can be used to measure 
$\beta_s$ without the need of an angular analysis 
as $J/\psi f_0(980)$ is a pure CP-odd final state. Further interest in the 
decay \bsjpsifzero\  arises from the fact  that it could contribute to an S-wave component in the  $B_s^0 \to J/\psi K^{+}K^{-}$ decay	
if $f_0(980)$ decays  to $K^{+}K^{-}$.	
CDF recently confirmed the observation of the \bsjpsifzero\  decay from the LHCb and Belle experiments~\cite{lhcb_belle_bsjpsifzero}.  
At present the observed signal is the world's largest and the measurement of the ratio
of branching fractions between $\bsjpsifzero$ and  \bsjpsiphi\  decays $0.257 \pm 0.020\stat \pm 0.014\syst$~\cite{cdf_bsjpsifzero} is the the most precise.
Using the world average \bsjpsiphi\ branching fraction \cite{pdg} this can be converted into the product of branching fractions of
$\mathcal{B}(\bsjpsifzero)  \mathcal{B}(f^{0}(980)\to \pi^+\pi^-)= (1.63 \pm 0.12 \pm 0.09 \pm 0.50) \times 10^{-4}$~\cite{cdf_bsjpsifzero}
where the first uncertainty is statistical, the second is systematic and the third one is due to the uncertainty 
on branching fraction of normalization channels.
The measurement presented here agrees well with the previous measurements of this quantity. CDF reports also the first determination of the lifetime 
of this decay mode $\tau=1.70^{+0.12}_{-0.11}\stat\pm 0.03 \syst$~\cite{cdf_bsjpsifzero}.

An alternative way of accessing $\phi_s$ is through the measurement 
of the asymmetry $A_{sl}^b$ defined as
$A^b_{sl} \equiv \frac{N^{++}_{b} - N^{--}_{b}}{N^{++}_{b} + N^{--}_{b}},$
where $N^{++}_{b}$ and $N^{--}_{b}$ represent the number of events
containing two $b$ hadrons decaying semileptonically and producing two positive or
two negative muons, respectively.  
Measurements of $A^b_{sl}$ or $\phi_q$ that differ
significantly from the SM expectations would indicate the presence of new physics.\\
D\O\ analysed a data sample corresponding to 
an integrated luminosity of $6.1$~fb$^{-1}$
and measured the like-sign dimuon charge asymmetry
of semileptonic $b$-hadron decays: $A^b_{sl} = -0.00957 \pm 0.00251~({\rm stat}) \pm 0.00146~({\rm syst})$~\cite{d0_asl},
which is in disagreement with the prediction of the standard model by
3.2 standard deviations, as reported in \fig{contours} by the 
band, which is a 68\% CL contour.

\section{\boldmath{$\gamma$} from \boldmath{$B \to D K$} decays}

Conventionally, CP violating observables are written in terms of the angles $\alpha$, $\beta$ and $\gamma$ of 
the ``Unitarity Triangle", obtained from one of the unitarity conditions of the CKM matrix. 
While the resolution on $\alpha$ and $\beta$ reached a good level of precision, the measurement of $\gamma$ is still limited 
by the smallness of the branching ratios involved in the processes. 
Among the various methods for the $\gamma$ measurement, those which make use of the tree-level $B^- \to D^0 K^-$ 
decays have the smallest theoretical uncertainties~\cite{ref:glw1,ref:ads1,ref:ggsz}. In fact $\gamma$ appears as the relative 
weak phase between two amplitudes, the favored $b \to c \bar{u} s$ transition of the $B^- \to D^0 K^-$, whose amplitude is 
proportional to $V_{cb} V_{us}$, and the color-suppressed $b \to u \bar{c} s$ transition of the $B^- \to \overline{D}^0 K^-$, 
whose amplitude is proportional to $V_{ub} V_{cs}$.  
The interference between $D^0$ and $\overline{D}^0$, decaying into the same final state, leads to measurable CP-violating 
effects, from which $\gamma$ can be extracted. The effects can be also enhanced choosing the interfering amplitudes of the same order of magnitude.
All methods require no tagging or time-dependent measurements, and many of them only involve charged particles in 
the final state. 
CDF reports the first results in hadron collisions for the ADS~\cite{ref:glw1} and GLW~\cite{ref:ads1} methods.

In a data sample of about 5~\lumifb\ CDF 
measures 
the following asymmetries:
$A_{ADS}(K)  =   - 0.63 \pm 0.40\mbox{(stat)} \pm 0.23\mbox{(syst)}$ and 
$A_{ADS}(\pi) =   0.22 \pm 0.18\mbox{(stat)} \pm 0.06\mbox{(syst)}.$~\cite{cdf_ads}.
For the ratios of doubly Cabibbo suppressed mode to flavor eigenstate CDF finds 
$R_{ADS}(K)  =  [22.5 \pm 8.4\mbox{(stat)} \pm 7.9\mbox{(syst)}] \cdot 10^{-3}$ and 
$R_{ADS}(\pi)  =  [4.1 \pm 0.8\mbox{(stat)} \pm 0.4\mbox{(syst)}] \cdot 10^{-3}$~\cite{cdf_ads}.
These quantities are measured for the first time in hadron collisions. The results
are in agreement with existing measurements performed at $\Upsilon$(4S) resonance~\cite{ads_bfactories}. 
The results on GLW method from CDF can be found in Ref.~\cite{cdf_glw}.


\section{Charm Physics}

While investigations of the $K$ and $B$ systems have and will
continue to play a role in the understanding of flavor physics and CP violation, 
the $D$ mesons sector have yet probed with a sufficient precision to explore the range of SM predictions. 
Since charm quark is the only up-type charged 
quark accessible to experiment though the $D$ mesons, 
it provides the unique opportunity to probe flavor physics in this sector that is complementary to the one of down-type quarks.
Examples of clean channels sensitive to possible sources of CP violation in the charm system are the singly-Cabibbo 
suppressed transitions such as $D^0\to\pi^+\pi^-$ and $D^0\to K^+ K^-$. 
Contribution to these decays from ``penguin'' amplitudes are negligible in the SM, thus the presence 
of new exotic particles may enhance the size of CP violation with respect to the SM expectation. 
Any asymmetry significantly larger than a few times $0.1\%$ may unambiguously indicate NP contributions. 

CDF recently measured the time-integrated CP asymmetry in the $D^0\to\pi^+\pi^-$ and $D^0\to K^+K^-$ decays using 
5.94 fb$^{-1}$ of data collected by the displaced track trigger. The final results, which are the most precise up to date, are
$A_{CP}(D^0\to\pi^+\pi^-) = \bigl[+0.22\pm0.24\stat\pm0.11\syst\bigr]\%\qquad$ and
$A_{CP}(D^0\to K^+K^-) = \bigl[-0.24\pm0.22\stat\pm0.10\syst\bigr]\%$~\cite{cdf_charm}. 
They are consistent with CP conservation and also with the SM predictions.
As expressed by Eq.~3 of Ref.~\cite{cdf_charm} the $A_{CP}(D^0 \to h^+h^-)$ measurement describes a 
straight line in the plane $(A_{CP}^{\rm ind},A_{CP}^{\rm dir})$ with angular coefficient given by $\langle t\rangle/\tau$,
being $t$ the measured proper decay time of the $D^0$ candidates and $\tau$ the lifetime of the $D^0$ meson.
Because of a threshold on the impact parameter of tracks, imposed at trigger level, the CDF 
sample of \mbox{$D^0\to\pi^+\pi^-$} (\mbox{$D^0\to K^+K^-$}) decays is enriched in 
higher-valued proper decay time candidates with a mean value of $2.40(2.65)\pm0.03\ (\mathit{stat.}+\mathit{syst.})$ 
times the $D^0$ lifetime, as measured from a fit to the proper time distribution. 
Due to their unbiased acceptance in charm decay time, $B$~factories samples have instead 
$\langle t\rangle\approx\tau$. Hence, the combination of the three 
measurements allow to constrain independently both $A_{CP}^{\rm dir}$ and $A_{CP}^{\rm ind}$~\cite{cdf_charm}. 





\end{document}

%% file: macro.tex
\def \rightdownarrow
 {\kern.3em
 \rule[.5ex]{.15mm}{2ex}
 {\mbox{$\kern-0.1em{\longrightarrow}$}}      
 }

\def\lessim{\mathrel {\vcenter {\baselineskip 0pt \kern 0pt  
\hbox{$<$} \kern 0pt \hbox{$\sim$} }}}

\def\gessim{\mathrel {\vcenter {\baselineskip 0pt \kern 0pt   
\hbox{$>$} \kern 0pt \hbox{$\sim$} }}}

\newcommand{\lumifb}{\mbox{fb$^{-1}$}}				

\newcommand{\br}{\ensuremath{\mathcal{B}}}

\newcommand{\tev}{\ensuremath{\mathrm{Te\kern -0.1em V}}}
\newcommand{\gev}{\ensuremath{\mathrm{Ge\kern -0.1em V}}}	
\newcommand{\mev}{\ensuremath{\mathrm{Me\kern -0.1em V}}}	
\newcommand{\kev}{\ensuremath{\mathrm{ke\kern -0.1em V}}}	

\newcommand{\stat}{\ensuremath{\mathit{~(stat.)}}}		
\newcommand{\syst}{\ensuremath{\mathit{~(syst.)}}}		

\newcommand{\CP}{CP}			

\newcommand{\bs}{\ensuremath{B^{0}_s}}				

\newcommand{\abs}{\ensuremath{\overline{B}^{0}_s}}		

\newcommand{\bsjpsiphi}{\ensuremath{\bs \to  \jpsi \phi}}

\newcommand{\bsjpsifzero}{\ensuremath{\bs \to  \jpsi f_0(980)}}

\newcommand{\jpsi}{\ensuremath{J/\psi}}

\newcommand{\fig}[1]{fig.~\ref{fig:#1}}

\newcommand{\cita}[1]{\cite{#1}}

\def\babar{\mbox{\slshape B\kern-0.1em{\smaller A}\kern-0.1em B\kern-0.1em{\smaller A\kern-0.2em R}}}